# Molecular determinants involved in the allosteric control of agonist affinity in GABA$_B$ receptor by the GABA$_{B2}$ subunit

Jianfeng Liu[‡&], Damien Maurel[‡+], Sébastien Etzol[‡], Isabelle Brabet[‡], Hervé Ansanay[+], Jean-Philippe Pin[‡] & Philippe Rondard[‡*]

[‡] Department of Molecular Pharmacology, Laboratory of Functional Genomic, CNRS UPR2580, CCIPE, Montpellier, France, and [+] Cis Bio International, Bagnols-sur-Cèze, France.

[&] Institute of Biophysics and Biochemistry, Huazhong University of Science and Technology, 1037 Luoyu Avenue, 430074 Wuhan, Hubei, China.

*Running title:* Allosteric communication between GABA$_B$ subunits

*To whom correspondence should be addressed:
   Dr. Philippe Rondard
   Department of Molecular Pharmacology, Laboratory of Functional Genomic, CNRS UPR2580, CCIPE, 141 Rue de la Cardonille, F-34094 Montpellier cedex 05, France.
   Phone: +33 4 67 14 29 12    Fax: +33 4 67 54 24 32
   Email: prondard@ccipe.cnrs.fr

This work was supported by grants from the CNRS, the Fondation pour la Recherche Médicale, and the French government (Action Concertée Incitative "Molécules et Cibles Thérapeutiques") to J.-P. P.. J.L. was funded from a grant from Aventis (GIP-HMR to J.-P. P.), and P.R by the Fondation pour la Recherche Médicale (France) and INSERM. D. M. was supported by Cis Bio International (Marcoule, France).

**Abbreviations:**

FRET, Fluorescence resonance energy transfer; GABA, γ-aminobutyrate; GB1, $GABA_{B1}$ subunit; GB2, $GABA_{B2}$ subunit; GPCR, G-protein coupled receptors; HA, hemagglutinin; HD, heptahelical domain; HEK-293 cells, human embryonic kidney cells; mGlu receptors, metabotropic glutamate receptors; TR-FRET, time resolved FRET; VFT, Venus Flytrap domain.

**Abstract**

γ-aminobutyric acid (GABA) type B receptor (GABA$_B$) is an allosteric complex made of two subunits, GB1 and GB2. Both subunits are composed of an extracellular Venus flytrap domain (VFT) and a heptahelical transmembrane domain (HD). GB1 binds GABA and GB2 plays a major role in G-protein activation as well as in the high agonist affinity state of GB1. How agonist affinity in GB1 is regulated in the receptor remains unknown. Here, we demonstrate that GB2 VFT is a major molecular determinant involved in this control. We show that isolated versions of GB1 and GB2 VFTs, in the absence of the HD and C-terminal tails, can form hetero-oligomers as shown by time-resolved FRET. GB2 VFT and its association with GB1 VFT control agonist affinity in GB1 in two ways. First, it exerts a direct action on GB1 VFT as GB2 VFT increases slightly agonist affinity on the isolated GB1 VFT. Second and most importantly, GB2 VFT prevents inhibitory interaction between the two main domains (VFT and HD) of the GB1 subunit. According to this model, we propose that the HD of GB1 prevents the possible natural closure of the GB1 VFT. In contrast, GB2 VFT facilitates this closure. Finally, such inhibitory contacts between HD and VFT in GB1 could be similar to those important to maintain the inactive state of the receptor.

# INTRODUCTION

γ-aminobutyric acid (GABA), a major inhibitory neurotransmitter in the central nervous system, activates $GABA_A$ ligand-gated $Cl^-$ channels, as well as the G-protein-coupled receptor (GPCR) $GABA_B$ (1,2). This receptor is found in either pre- or post-synaptic elements in various types of neurons. As such, $GABA_B$ receptors play important role in brain function as illustrated by the antispastic activity of the $GABA_B$ selective agonist baclofen (Lioresal®), and its involvement in various types of epilepsy, as well as in nociception and drug addiction (2).

$GABA_B$ receptors belong to the class-III GPCRs, together with metabotropic glutamate (mGlu), extracellular $Ca^{2+}$-sensing, and some pheromone and taste receptors (3). Each of these receptors is composed of an extracellular domain called Venus flytrap (VFT) where agonists bind, and an heptahelical domain (HD) responsible for the recognition and activation of heterotrimeric G-proteins. Whereas mGlu and $Ca^{2+}$-sensing receptors exist as homodimers, $GABA_B$ receptor is an heterodimer composed of two homologous subunits, GB1 and GB2 (4-7) (Fig. 1a).

So far, only the heterodimeric form of the $GABA_B$ receptor has been shown to activate G-proteins efficiently. Although only GB1 binds GABA, several important roles of GB2 have been identified. First, GB2 masks an intracellular retention signal of GB1, such that GB1 reaches the cell surface only when associated with GB2 (2,8-10). Second, the HD of GB2 contains all the determinants required for G-protein coupling and plays a pivotal role in G-protein activation by the heteromer (11-15). Third, GB2 increases agonist, but not antagonist affinity on GB1 (4,11), even though it does not appear to bind any natural ligand (16).

The ligand binding site of GB1 has been extensively studied (17-22). Modeling and site-directed mutagenesis studies indicate ligands bind in the cleft that separates both lobes of the GB1 VFT, as observed for ligand binding in many similar protein modules (23) including the mGlu1 VFT (24,25). Antagonists are expected to prevent the closure of the GB1 VFT (19), as observed in mGlu receptors (25,26). Conversely, agonists interact with residues from both lobes of GB1 VFT and they stabilize a closed form of this domain (16,19). Such a domain closure of the GB1 VFT has recently been shown to be sufficient to activate this heterodimeric receptor (27).

Mechanism of the allosteric control of agonist affinity in GB1 by GB2 is unknown. It is likely that the GB2 subunit controls agonist affinity by further stabilizing the closed state of the GB1 VFT (28). Understanding this mechanism should have implications in the current model of $GABA_B$ receptor activation, as it may help to explain how GABA binding in the GB1 VFT can activate the GB2 HD. In addition, it may open new routes for the development of positive allosteric compounds known to stabilize the active conformation of the dimeric HDs and VFTs and to increase agonist affinity (29-31).

Here, we demonstrate that direct interaction between GB1 and GB2 VFTs is responsible for the increase in agonist affinity in two ways. First, it prevents the HD of GB1 to decrease agonist affinity, and second, the interaction between the two VFTs by itself further increases agonist affinity in GB1 VFT.

# MATERIAL AND METHODS

**Materials**

γ-aminobutyric acid (GABA) was obtained from Sigma (Saint-Quentin Fallavier, France). CGP64213 was gift from Drs. W. Froestl and K. Kaupmann (Novartis Pharma, Basel, Switzerland). [$^{125}$I]-CGP64213 was purchased from Anawa (Zurich, Switzerland). Culture media, foetal bovine serum (FBS) and other solutions used for cell culture were from Invitrogen (Cergy Pontoise, France).

**Contruction of GB1 and GB2 mutants**

The plasmids pRK5 encoding the wild-type GB1a, GB1$_{ASA}$ and GB2 subunits tagged with the hemagglutinin (HA) or c-myc epitope at their N-terminal ends, under the control of a cytomegalovirus promotor, were described previously (10,11). Truncated versions of GB1 and GB2 (ΔGB1 and ΔGB2) were generated by replacing codons for Ile621 of GB1 and Leu511 of GB2 by a stop codon using Quik-Change mutagenesis protocol (Stratagene, La Jolla, CA). Glycosylphosphatidylinositol-anchored VFT module of GB1 (GB1$_{GPI}$) was constructed by subcloning of a synthetic gene fragment incoding for the GPI anchor signal peptide of the mouse PrPc (32), after the entire coding sequence of GB1 VFT (*MluI-PshAI* fragment). *PshAI-XbaI* fragment of pRK-GB1a-HA, encoding for HD and C-terminus domains of GB1, was replaced by a synthetic gene fragment created by hybridization of two complementary oligonucleotides 5'-TGG TCA GAA GAT CCA GCA GCA CCG TGC TTT CCT CCC

CTC CTG TCA TCC TCC TCA TCT CCT TCC TCA TCT TCC TGA TCG TGG GAT AAT-3' and 5'-CTA GAT TAT CCC ACG ATC AGG AAG ATG AGG AAG GAG ATG AGG AGG ATG ACA GGA GGG GAG GAG AAA AGC ACG GTG CTG CTG GAT CTT CTG ACCA-3'. It results in the addition of the amino acid sequence RRS<u>S</u>STVLFSSPPVILLISFLIFLIVG after residue Val579 of GB1. Post-translationally, GPI anchor signal peptide is cleaved after the second serine residue (S) of the tripeptide SSS (printed in underlined) and a GPI modification is added on that one (33). Glycosylphosphatidylinositol-anchored VFT module of GB2 (GB2$_{GPI}$) was generated by PCR and subcloned into the plasmid pRK5-GB1$_{GPI}$, in replacement of *MluI-PshAI* fragment encoding for GB1 VFT. In GB2$_{GPI}$, GPI anchor signal peptide is added after residue Gln473 of GB2 following by Thr-Leu-Val sequence due to the *PshAI* site. A truncated version of metabotropic mGlu5 receptor (ΔmG5) in the first intracellular loop (after residue Arg614) was generated by PCR from mGlu5 receptor expression vector previously described (34), and subcloned into pRK5-GB1-HA or -c-myc, in replacement of *MluI-XbaI* fragment encoding for the entire sequence of GB1. Mutation S246A was introduced in ΔGB1 by subcloning *ApaI-PshAI* fragment of pRK5-GB1(S246A) (19) into pRK5-ΔGB1.

**Cell culture and transfection**

HEK-293 and COS-7 cells were cultured in Dulbecco's modified Eagle's medium supplemented with 10% FBS and transfected by electroporation as described elsewhere (35).

Ten million cells were transfected with plasmid DNA containing GB1$_{ASA}$ (2 μg), GB2 (2 μg), ΔGB1 (4 μg), ΔGB2 (4 μg) , GB1$_{GPI}$ (4 μg), GB2$_{GPI}$ (4 μg), ΔmG5 (4 μg), ΔGB1(S246A) (4 μg) and V2 vasopressin receptor (165 ng) and completed to a total amount of 10 μg plasmid DNA with pRK5 empty vector.

**Western blotting**

Twenty hours after transfection, HEK-293 cells were washed with PBS (Ca$^{2+}$- and Mg$^{2+}$- free) and harvested. The membranes were prepared as previously described (26). For each sample, 50 μg of total protein was subjected to SDS-PAGE by using 10 % polyacrylamide gels, transferred to nitrocellulose membrane (Hybond-C; Amersham Pharmacia), and probed with anti-HA mouse monoclonal antibody (clone 12CA5; Roche, Basel, Switzerland) at 0.1 μg/mL. Proteins were visualized by chemiluminescence (West Pico; Pierce, Rockford, IL).

**Cell surface quantification by ELISA**

Twenty hours after transfection with HA-tagged versions of the constructs, HEK-293 cells were washed twice with phosphate-buffered saline solution (PBS), fixed with 4% paraformaldehyde in PBS and then blocked with PBS plus 5% FBS. After 30 minutes reaction with primary antibody (monoclonal anti-HA clone 3F10; Roche, Basel, Switzerland) at 0.5 μg/mL) in the same buffer, the goat anti-rat antibody coupled to horseradish peroxidase (Jackson Immunoresearch, West Grove, PA) was applied for 30 minutes at 1 μg/mL. After intense washes with PBS, secondary antibody was detected and quantified instantaneously by chemiluminescence (Supersignal West Femto; Pierce, Rockford, IL) using a Wallac Victor$^2$ luminescence counter (Molecular Devices, St Grégoire, France).

**Ligand binding assay**

Ligand binding assay on intact HEK-293 cells was performed as previously described using 0.1 nM [$^{125}$I]-CGP64213 (11). Displacement curves were performed with at least 7 different concentrations of the displacer, and the curves were fitted according to the equation: "$y=[(y_{max}-y_{min})/(1+(x/IC_{50})^{nH})] +y_{min}$" where the $IC_{50}$ is the concentration of the compound that inhibits 50% of bound radioligand and nH is the Hill coefficient. Ki values were calculated according to the equation $IC_{50}=Ki(1+[RL]/Kd)$, where [RL] and Kd are the concentration and dissociation constant of the radioligand. Kd was determined assuming Ki=Kd in the case of CGP64213.

**Time-resolved FRET measurements**

Time-resolved FRET experiments were conducted as described by Maurel et al.[1]. After transfection, COS-7 cells were dispatched into a black 96-wells assay plate (COSTAR) at 1.5 x 10$^5$ cells per well in 200 μl of Dulbecco's modified Eagle's medium supplemented with 10% FBS. Twenty hours later, cells were rinsed in 100 μl Tris-Krebs buffer (20 mM Tris-Cl, pH 7.4, 118 mM NaCl, 5.6 mM glucose, 1.2 mM KH$_2$PO$_4$, 1.2 mM MgSO$_4$, 4.7 mM KCl, 1.8 mM CaCl$_2$) supplemented with BSA 0.1%. Then, cells were incubated in 100 µl of the same buffer containing 3 nM of anti-HA monoclonal antibodies (12CA5) labeled with a fluorescence donor molecule (europium cryptate-PBP[1]) and 9 nM of anti-c-myc monoclonal antibodies (9E10) labeled with a fluorescence acceptor molecule (Alexa Fluor® 647; Molecular Probes) (these two antibodies were kind gifts by Cis Bio International, Marcoule, France). As negative control, 1 µM of unlabeled anti-c-myc antibodies (9E10) was added to the couple of labeled antibodies in order to displace the FRET signal, or COS-7 were incubated with only fluorescence donor antibodies. After four hours incubation at 4°C, cells were rinsed with Tris-Krebs buffer to remove unbound antibodies, and fluorescence emissions were monitored both at 620 nm and at 665 nm with a RubyStar spectrofluorimeter equipped with a nitrogen laser as excitation source (337 nm) (BMG LabTechnologies, Champigny-sur-Marne, France). A 400-µsec reading was used after a 50-µsec delay to allow for decay of short-lived endogenous fluorescence signals. Fluorescence collected at 620 nm is the total europium cryptate signal and fluorescence at 665 nm is the FRET signal, and the ratio R=[(fluorescence 665 nm/ fluorescence 620 nm) x 10$^4$] was computed. FRET signal was expressed by Delta F (ΔF) calculated using the formula ΔF = (Rpos - Rneg)/(Rneg) where Rpos is the ratio for the positive energy transfer measured in wells incubated with both donor and acceptor antibodies, and Rneg is the ratio for the negative energy transfer control. Total fluorescence emitted at 682 nm by the Alexa Fluor® 647 conjugates after excitation at 640 nm was measured using an Analyst™ reader (Molecular Devices) equipped with the appropriate filters set (XF47 from Omega Optical).

---

[1] Maurel et al., manuscript in preparation

# RESULTS

**GB2 increases agonist affinity, but not antagonist affinity on GB1**

Although GB1 bind any known $GABA_B$ ligands, agonist affinities are about 100 times lower than those measured on native receptors (36). This is partly due to GB2, since its co-expression with GB1 results in a 10 fold increase in agonist affinity (4). This effect of GB2 does not result from the targeting of GB1 to the cell surface, and so to a mature glycosylation state, since a GB1 mutant able to reach the cell surface alone ($GB1_{ASA}$ in which the ER retention signal RSRR is mutated into ASAR) still displays a low agonist affinity at the cell surface (Fig. 1b and Table 1). A direct association between GB1 and GB2 appears therefore necessary to control agonist affinity in this receptor. Co-expression of GB1 with GB2/1, a chimeric subunit composed of the VFT of GB2 and the HD of GB1, or the replacement of the HD of GB1 by that of GB2 in the chimeric GB1/2 subunit, also resulted in an increased GABA affinity (Fig. 1b and Table 1). These observations suggest that both the GB2 VFT and the HD of GB1 control agonist affinity on GB1.

**VFT module constructs for GB1 and GB2**

In order to elucidate the mechanism leading both GB2 VFT and GB1 HD to allosterically control agonist affinity on GB1, isolated GB1 and GB2 VFTs were prepared by removing the HD and C-terminal domains of each subunit. These constructs were generated to examine the possible direct interaction between GB1 and GB2 VFTs and the consequence on GABA affinity, regardless the other regions of the subunits. To express either GB1 or GB2 VFTs at the cell surface, two series of constructs were prepared. The first constructs called ΔGB1 and ΔGB2 were generated by introducing a stop codon at the end of the first intracellular loop of GB1 and GB2, respectively. As such, the VFTs were anchored in the plasma membrane by the first transmembrane helix (TM1) of these subunits (Fig. 2a). For the second constructs called $GB1_{GPI}$ and $GB2_{GPI}$, the entire HD and C-terminal tail were replaced by the glycosylphosphatidylinositol (GPI) anchor sequence of the PrPc prion protein (see Material and Methods and Fig. 2a). In all cases, these constructs contain an N-terminal epitope, either c-myc or HA inserted after a signal peptide allowing their easy detection at the cell surface. Previous studies demonstrated that such epitope affected neither the pharmacology nor the function of these subunits (10,11).

All constructs are correctly expressed in HEK-293 cells, and display the expected molecular weight as shown in Western blot experiments (Fig. 2b). All constructs except $GB2_{GPI}$ were found at the cell surface but at a density two to ten times lower than that of $GB1_{ASA}$ as revealed by an anti-HA ELISA performed on intact cells (Fig. 2c). Finally, ΔGB1 and $GB1_{GPI}$ bind a competitive and membrane non-permeant radio-labeled antagonist [$^{125}$I]CGP64213, and this binding can be displaced by GABA demonstrating these constructs retained their ability to bind $GABA_B$ ligands (Fig. 2d and Table 1). All together, these results show that ΔGB1, ΔGB2 and GB1$_{GPI}$ are correctly expressed at the cell surface, and that ΔGB1 and GB1$_{GPI}$ are correctly folded.

To examine whether an interaction between GB1 and GB2 VFTs could be detected, the above described truncated GB1 and GB2 subunits as well as the wild-type subunits were co-expressed. GB2 and ΔGB2 increase expression of ΔGB1 and GB1$_{GPI}$ at the cell surface. Binding experiments show that total amount of radioligand bound to ΔGB1 at the cell surface is increased in the presence of ΔGB2 (Fig. 3a). A higher increase is observed when ΔGB1 is co-expressed with the full-length GB2 (Fig. 3a). Similarly, the amount of [$^{125}$I]CGP64213 bound to GB1$_{GPI}$ at the cellular surface is also increased when this construct is co-expressed with either ΔGB2 and GB2 (Fig. 3a). Since neither GB2 nor ΔGB2 changes CGP64213 affinity on these GB1 constructs (Table 1), these data show that GB2 and ΔGB2 increase the number of [$^{125}$I]CGP64213 binding sites at the cell surface. We further confirm that GB2 and ΔGB2 increase the amount of truncated GB1 constructs at the cell surface using an ELISA on intact cells. The amount of HA-ΔGB1 at the cell surface is increased after co-expression with c-myc-ΔGB2 or c-myc-GB2 (Fig. 3b). Thus, GB2 VFT either stabilizes ΔGB1 and GB1$_{GPI}$ at the cell surface, or facilitates their targeting to the plasma membrane.

**GB1 and GB2 VFT modules assemble into heterodimeric complexes in the absence of heptahelical domain**

To demonstrate more directly that GB1 and GB2 VFTs interact with each other, co-immunoprecipitation experiments were performed. Unfortunately, no interaction between GB1 and GB2 VFTs was detected (data not shown), possibly because such complexes are not stable enough and did not resist to the sample preparation. Indeed, although the deletion of the C-terminal coiled-coil domain of the GB1 and GB2 subunits does not prevent heterodimer formation, as shown by the normal functioning of the receptor, it largely decreases the amount of GB2 co-precipitated with GB1 (9). This shows that the coil-coiled domains in the C-terminal tails of these subunits strongly stabilize the heterodimer.

We therefore conducted time-resolved fluorescence resonance energy transfer (TR-FRET) experiments as previously described (37) ( and Maurel et al., submitted). In this assay, a FRET signal is measured at the surface of intact COS-7 cells between a donor molecule (europium cryptate-PBP) linked to an anti-HA monoclonal antibody, and an acceptor molecule (Alexa Fluor® 647) linked to an anti-c-myc-monoclonal antibody (Fig. 4a, Insert). In this assay, the HA-tagged version of $\Delta$GB1 and GB1$_{GPI}$ and the c-myc tagged version of $\Delta$GB2 and GB2 (or other control constructs) were used. As shown in Fig. 4a and 5a, a FRET signal is measured at 665 nm (the emission wavelength of Alexa Fluor® 647) after excitation of the europium cryptate-PBP at 337 nm in cells expressing either HA-$\Delta$GB1 or HA-GB1$_{GPI}$ together with c-myc-ΔGB2. To better compare the association efficiency of the different partners, FRET signal was normalized with regard to c-myc tagged construct expression (4c et 5c). This signal is significantly higher than that measured in mock transfected cells. Moreover, such a FRET signal was not detected when ΔGB1 and GB1$_{GPI}$ were co-expressed with the c-myc-tagged V2 vasopressin receptor, a class-I GPCR (Fig. 4a and 5a). As shown in Fig. 4b and 5b, this did not result from the low expression at the cell surface of the c-myc tagged partners. A significant FRET signal was however measured between ΔGB1 and ΔmG5 (the VFT domain of metabotropic mGlu5 receptor anchored to in the plasma membrane via the first TM) and to a lower extent between ΔGB1 and V2-R (Fig. 4a). However, such signals were not observed with GB1$_{GPI}$ (Fig. 5a) suggesting that TM1 of ΔGB1 is likely involved in a non-specific interaction with other TM containing proteins. This may be because the TM used to anchor ΔGB1 at the cell surface (TM1 of GB1) is usually associated with other TMs in the GB1 HD.

During these experiments, we also noticed GB1 and GB2 VFT modules form homomeric complexes, as previously reported for both full-length GB1 and GB2 subunits [1]. Indeed, a FRET signal could be measured in cells expressing HA-GB1$_{GPI}$ and c-myc-ΔGB1. This signal is similar to that obtained with HA-ΔGB1 and c-myc-ΔGB2 or HA-GB1$_{GPI}$ and c-myc-ΔGB2 (Fig. 4a and 5a), consistent with GB1 VFT being able to oligomerize. Similar FRET signal between ΔGB1 expressed alone and GB1$_{GPI}$ co-expressed with ΔGB1 suggest

TM1 of ΔGB1 is not responsible for its homomerisation. Moreover, we found that the GB2 VFT can also form homo-oligomers in similar FRET experiments (data not shown).

**Association between VFT modules of GB1 and GB2 increases agonist affinity on GB1**

We then examined the possible influence of the GB1-GB2 VFTs interaction on agonist affinity. To that aim GABA affinity was measured by displacement of radioligand [$^{125}$I]CGP64213 on intact cells expressing the above described constructs alone or in combination (Table 1). In contrast to agonist, we verified that antagonist CGP64213 displaced with a similar potency the radioligand for every combination examined, indicating that none of the constructs co-expressed with GB1 affect antagonist affinity (Table 1).

As shown in Fig. 6a, ΔGB2 increases by a factor 16 GABA affinity on GB1$_{ASA}$ (Ki values were 16.0 ± 1.4 and 1.0 ± 0.2 μM for GB1$_{ASA}$ expressed alone or with ΔGB2 respectively). This effect is specific since no change in GABA affinity was observed when GB1$_{ASA}$ was co-expressed with ΔmG5 or ΔGB1(S246A), a GB1 construct unable to bind GABA$_B$ ligands (19). Association between GB1$_{ASA}$ and ΔGB2 or ΔGB1(S246A) was confirmed by measuring FRET signal between the co-expressed constructs (Fig. 6a, Insert). These results further confirm ΔGB2 associates with GB1$_{ASA}$ and show that this interaction increases GABA affinity on GB1.

Of interest, and as previously reported (22), GABA affinity on the isolated VFT of GB1 (either ΔGB1 or GB1$_{GPI}$) is close to that measured when GB1 is co-expressed with GB2 (Ki values 1.0 ± 0.2, 0.7 ± 0.1 and 3.2 ± 0.2 µM, respectively), and 10-20 times higher than that on GB1$_{ASA}$ (Table 1 and Fig. 6). This indicates that the HD of GB1 exerts an inhibitory action on agonist affinity. GABA affinity on these truncated GB1 constructs can still be further increased, although to a lower extent, after co-expression with either GB2 or ΔGB2 (3 and 2.5 fold, respectively) (Fig. 6b). Indeed, Ki values for GABA on ΔGB1 decreases from 1.0 ± 0.2 $\mu$M to 0.3 ± 0.1 $\mu$M and 0.4 ± 0.2 in the presence of GB2 and ΔGB2, respectively. This shows that most of the increased agonist affinity resulting from GB1-GB2 association is due to GB2 VFT preventing the HD of GB1 to decrease agonist affinity, rather than a direct effect of GB2 VFT on the GB1 VFT closure.

# DISCUSSION

In this study, we show how the association between the VFTs of $GABA_B$ receptor subunits GB1 and GB2 can allostericaly influence agonist affinity. GB2 VFT controls GB1 affinity for GABA in two ways. First, it exerts a direct action on GB1 VFT as GB2 VFT increases slightly agonist affinity on the isolated GB1 VFT. Second and most importantly, GB2 VFT prevents inhibitory interaction between the two main domains (VFT and HD) of the GB1 subunit.

**Negative allostery within the GB1 subunit**

Inhibitory contacts that maintain GB1 subunit alone in a low affinity state may have several origins. One possibility is that the direct interaction between the HD and the VFT domain within the GB1 subunit contrains the VFT in a low agonist affinity state. Alternatively, GB1 can exist as homodimer and interactions between the GB1 VFTs in the homodimer could stabilize a low agonist affinity state. We exclude this latter possibility since the isolated VFT of GB1 (both ΔGB1 and $GB1_{GPI}$) adopts a high agonist affinity state even though it is able to homodimerize. Accordingly, a direct interaction between the HD and the VFT domains of GB1 is likely responsible for the low agonist affinity state. Such an interaction has already been proposed by others but has never been demonstrated (22,38).

Such a negative effect of the HD on agonist affinity in the VFT has also been reported for other class-III GPCRs, the mGlu4 and mGlu8 receptors, using a similar approach (39,40).

Negative allostery of GB1 is controlled by GB2 VFT, and such a process is probably not a specificity of the expression of this subunit in heterologous cells. Indeed, only a high agonist affinity GABAB receptor is found in the brain. Moreover, GB1 needs to be associated with GB2 to reach the cell surface in the central nervous system, and neither GB1-GB1 nor GB2-GB2 homodimers couple efficiently to G-proteins (4,6).

**How is the agonist affinity controlled?**

Both GB1 and GB2 VFTs are structurally related to the VFT of mGlu1 metabotropic receptor and to bacterial periplasmic binding proteins, as suggested by modeling studies (16,17,19,21). Such domains are well-known to adopt either an open conformation (VFTo) stabilized by antagonists (25,26), or a closed conformation (VFTc) stabilized by agonists (24,26,27). VFT can oscillate between VFTo and VFTc states with an equilibrium constant $K1 = [VFTc]/[VFTo]$. A bound ligand will affect K1 by a factor $\alpha > 1$ in the case of an agonist that stabilizes VFTc, and $\alpha < 1$ in the case of an antagonist that stabilizes VFTo or prevents the VFT to reach the VFTc state (28).

Accordingly, ligand affinity in a VFT (Kd) depends on both the affinity of the ligand in VFTo ($K_L$), K1 and $\alpha$:

$$Kd = K_L (1+K1)/(1+\alpha K1)$$

According to this model, an increase in K1 (the receptor has a better tendency to reach the closed state in the absence of ligand) results in a large increase in agonist affinity (28). Moreover, and as observed in the case of the $GABA_B$ receptor, changing K1 results in minor changes in antagonist affinity (28). We therefore propose that in the absence of GB2, HD of GB1 decreases K1 (favors the VFTo state), an effect that is prevented when GB2 VFT associates with GB1 VFT. In addition, GB1-GB2 VFTs association may further increase K1 (favors the VFTc state) and thus agonist affinity.

**Implications for receptor activation**

GB2 VFT and its association with GB1 VFT appear to play a crucial role in $GABA_B$ receptor activation. In the absence of GABA, GB2 VFT constrains the receptor in an inactive state whereas in the presence of agonist, it facilitates receptor activation. Indeed, the presence of both GB1 and GB2 VFTs in the dimeric receptor is necessary for agonist stimulation. A dimeric receptor constituted by GB1-GB2 HDs but possessing two identical VFTs (either GB1 or GB2) display a large constitutive activity that cannot be further stimulated by agonists (11).

Inhibitory contacts between HD and VFT in GB1 that controls agonist affinity could be similar to those important for $GABA_B$ receptor activation. This model is supported by several pieces of data. First, the fact that GB2 VFT increases agonist but not antagonist affinity suggests that GB2 VFT stabilizes the active state of GB1. Second, we showed GB2

VFT acts by releasing inhibitory contacts between HD and VFT of GB1. Third, positive allosteric compounds of $GABA_B$ that likely bind in the HD increase both affinity and efficacy of agonists (29). Finally, similar inhibitory interactions between HD and VFT in GB2 subunit could exist due to the allosteric nature of the $GABA_B$ receptor, and these contacts could also play a role in receptor activation.

**Conclusion**

In conclusion, our experiments emphasize the functional importance of VFT interaction within the $GABA_B$ heterodimeric receptor and potentially to other class-III GPCRs. Our data show that the direct interaction between the VFTs is not only important for agonist activation of the receptor, as already reported for both $GABA_B$ (11) and mGlu receptors (41), but also for the control of agonist affinity. Indeed, a recent study show that mutations in the mGlu1 VFT that possibly prevents direct VFTs interaction within the receptor dimer largely decrease agonist affinity (41). Perspectives of this work is to identify regions of VFT and HD responsible for inhibitory contacts both at the GB1-GB2 VFTs and HD-VFT interfaces.

**Acknowledgements** – The authors wish to thank L. Prézeau, T. Durroux, F. Rassendren, J. Kniazeff and V. Perrier for helpful discussion, T. Durroux for the generous gift of the tagged version of the V2 receptor, and K. Kaupmann and W. Froestl to allowed us to use the radioligand [$^{125}$I]CGP64213. The authors also express their special thanks to Drs G. Mathis and E. Trinquet from Cis Bio International for their strong support.

|  | GABA Ki (μM) | CGP64213 Ki (nM) |
|---|---|---|
| **GB1** | 22.3 ± 1.9* | 1.4 ± 0.2* |
| **GB1/2** | 8.2 ± 0.9° | 4.0 ± 1.2° |
| **GB1 + GB2** | 3.3 ± 0.6° | 2.9 ± 0.2° |
| **GB1$_{ASA}$** | 16.0 ± 1.4 | 2.5 ± 0.2 |
| **GB1$_{ASA}$ + GB2** | 3.2 ± 0.3 | 2.8 ± 0.3 |
| **GB1$_{ASA}$ + GB2/1$_{ASA}$** | 1.2 ± 0.4° | 3.6 ± 0.6° |
| **GB1$_{ASA}$ + ΔGB2** | 1.0 ± 0.1 | 2.1 ± 0.6 |
| **GB1$_{ASA}$ + ΔmG5** | 14.2 ± 1.4 | 1.8 ± 0.3 |
| **GB1$_{ASA}$ + ΔGB1$_{S246A}$** | 18.3 ± 0.2 | 2.0 ± 0.1 |
| **ΔGB1** | 1.0 ± 0.2 | 2.1 ± 0.4 |
| **ΔGB1 + GB2** | 0.3 ± 0.1 | 1.5 ± 0.3 |
| **ΔGB1 + ΔGB2** | 0.4 ± 0.1 | 1.8 ± 0.2 |
| **GB1$_{GPI}$** | 0.7 ± 0.1 | 2.5 ± 0.3 |

**Table 1: Affinity values of GABA and CGP64213 for GB1 and ΔGB1 alone or in combination with different constructs**. Ki for GABA and CGP64213 were determined from displacement of [$^{125}$I]CGP64213 binding on intact cells expressing the indicated subunit combinations, as described in Materials and Methods. * indicates that the binding experiments

were performed on crude membranes (data from (19)). ° values taken from (11). Values are means ± s.e.m. of at least three independent determinations.

**Legends to Figures**

**Figure 1: VFT module of GB2 and HD of either GB1 and GB2 control agonist affinity on GB1.** (a) Schemes depict a dimeric $GABA_B$ receptor according to our actual view of the "resting" *(Left)* and "active" state *(Right)*. GB1 and GB2 subunits are in *dark gray* and *white*, respectively. In their "resting" orientation, both VFT modules in the dimer are in the open state, as observed in the crystal structure of the empty form of metabotropic receptor mGlu1 (PDB ID code 1ewt). One VFT is in the front plane (*dark gray*), whereas the other is in the back plane (*white*). The axis for the rotation of one VFT relative to the other and is indicated with a black circle, and the axis in each VFT responsible for their closure is indicated with an open circle. GABA (*black box*) binds in the cleft that separates both lobes of GB1 VFT and induces its closure leading to the active state. Each HD of GB1 and GB2 is composed of seven transmembrane helices (*oval boxes*) connected by extra- and intra-cellular loops, and C-terminus regions of the two subunits are associated (*dark gray lines*). *Asterisk* shows lobe-II of GB1 which is believed to interact with HD of GB1 and be responsible of the inhibitory contact inside GB1. (b) Displacement curves of [$^{125}$I]CGP64213 with GABA on cells

expressing GB1$_{ASA}$ (○), GB1 + GB2 (●) and GB1 + GB2/1 (□), where GB2/1 is a chimeric subunit composed of the VFT of GB2 and the HD of GB1. Percentage of bound radioligand for each concentration of GABA is plotted. Inhibitory concentration of GABA corresponding to the displacement of fifty percent of the initial bound radioligand (IC$_{50}$) is plotted for the three receptors. Values are means ± s.e.m. of at least three experiments performed in triplicates.

**Figure 2: VFT module constructs for GB1 and GB2** (a) Schemes of each individual constructs. VFT of GB1 (*dark gray*) and GB2 (*white*) were expressed with TM1 segment of the HD domain of GB1 (dark *gray oval*) and GB2 (*white oval*), respectively, or with a GPI anchor sequence (*broken lines*). VFT of mGlu5 (*black*) called ΔmG5 and a VFT of GB1 harboring a mutation S246A were expressed using TM1 segment of their own receptor. (b) Western blot analysis of the GB1 and GB2 constructs. The arrows indicate bands corresponding to ΔGB1 (lane 1), ΔGB2 (lanes 2), GB1$_{GPI}$ (lane 4) and GB2$_{GPI}$ (lane 5) and they are absent in mock-transfected cells (lanes 3 and 6). Constructs have an apparent molecular weight close to 60 kDa, as expected. (c) Cell surface expression of each individual HA-tagged constructs was measured by ELISA. Values are means ± s.e.m. of triplicates from a typical experiment and they are shown in arbitrary unit (A.U.). Although total expression of GB2$_{GPI}$ is similar to that of GB1$_{GPI}$, GB2$_{GPI}$ failed to reach the cell surface. (d) In *white*, binding of [$^{125}$I]CGP64213 for ΔGB1 and GB1$_{GPI}$ constructs expressed as a percentage of GB1$_{ASA}$. Specific binding is shown by the radioligand displacement by 1 mM GABA (*black bars*). Data are means ± s.e.m. of at least three experiments performed in triplicates.

**Figure 3: Cell surface expression of ΔGB1 and GB1$_{GPI}$ is increased by ΔGB2 and GB2.**

(a) Binding of radioligand on ΔGB1 and GB1$_{GPI}$ expressed alone or in combination with ΔGB2 or GB2. Data are means ± s.e.m. of at least three experiments performed in triplicates, and are expressed as percentage of the specific binding measured in cells expressing GB1$_{ASA}$. (b) Amount at the cell surface of ΔGB1 expressed alone or in combination with ΔGB2, GB2 or GB1$_{ASA}$ as measured by ELISA. Values are means ± s.e.m. of triplicates from a typical experiment and they are shown in arbitrary unit (A.U.)

**Figure 4: Association between VFT modules of ΔGB1 and ΔGB2 in time-resolved FRET experiments.** (a) Amount of FRET signal between co-expressed HA-tagged ΔGB1 and c-myc-tagged constructs. FRET signals are shown as ΔF (*See Materials and Methods*). *In insert*, scheme depicts the experimental system where VFT modules of GB1 (*dark gray*) and GB2 (*white*) are bound to a donor molecule (europium cryptate-PBP) linked to an anti-HA monoclonal antibody and to an acceptor molecule (Alexa Fluor® 647) linked to an anti-c-myc monoclonal antibody, respectively. FRET signal between the two antibodies is measured at 665 nm (E665) after excitation at 337 nm. (b) Amount of c-myc tagged constructs as measured by the fluorescence of Alexa Fluor® 647 antibody at 682 nm (E682). Signal intensities are normalized relative to GB2. Values are means ± s.e.m. of triplicates from a typical experiment. (c) ΔF FRET signal was normalized by the amount of c-myc-tagged constructs expressed at the cell surface. Experiments were done at least three times.

**Figure 5: GB1$_{GPI}$ and ΔGB2 interact together specifically in time-resolved FRET experiments.** (a) Amount of FRET signal ΔF between co-expressed HA-tagged GB1$_{GPI}$ and c-myc-tagged constructs, similarly to *Figure 4*. (b) Amount of c-myc tagged constructs as measured by the fluorescence emission of Alexa Fluor® 647 antibody (E682). Signal intensities are normalized relative to GB2. (c) ΔF FRET signal was normalized by the amount of c-myc-tagged constructs expressed at the cell surface. Experiments were done at least three times.

**Figure 6: Association between VFT modules of GB1 and GB2 increases agonist affinity on GB1.** (a) Displacement curves of [$^{125}$I]CGP64213 with GABA on cells expressing GB1$_{ASA}$ alone (○) or in combination with ΔGB2 (■), ΔmG5 (●) or ΔGB1(S246A) (❏), as a percentage of bound radioligand (*upper panel*). IC$_{50}$ were plotted from the displacement curve (*lower panel*). Values are means ± s.e.m. of at least three experiments performed in triplicates. In *upper panel insert*, association between HA-tagged GB1$_{ASA}$ and c-myc-tagged ΔGB2, ΔmG5 or ΔGB1(S246A) constructs as measured in time-resolved FRET experiments. (b) Displacement curves of the radioligand with GABA on cells expressing ΔGB1 alone (●) or in combination with ΔGB2 (■) or GB2 (❏), and corresponding GABA IC$_{50}$.

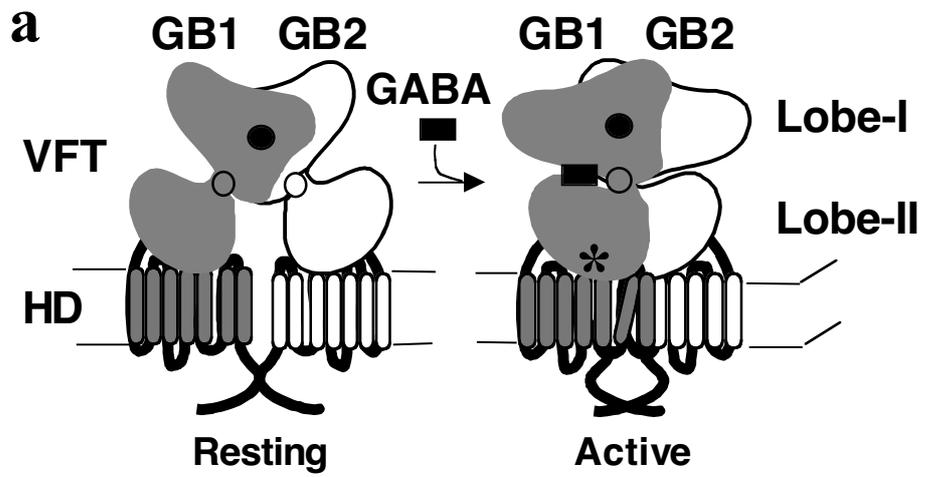
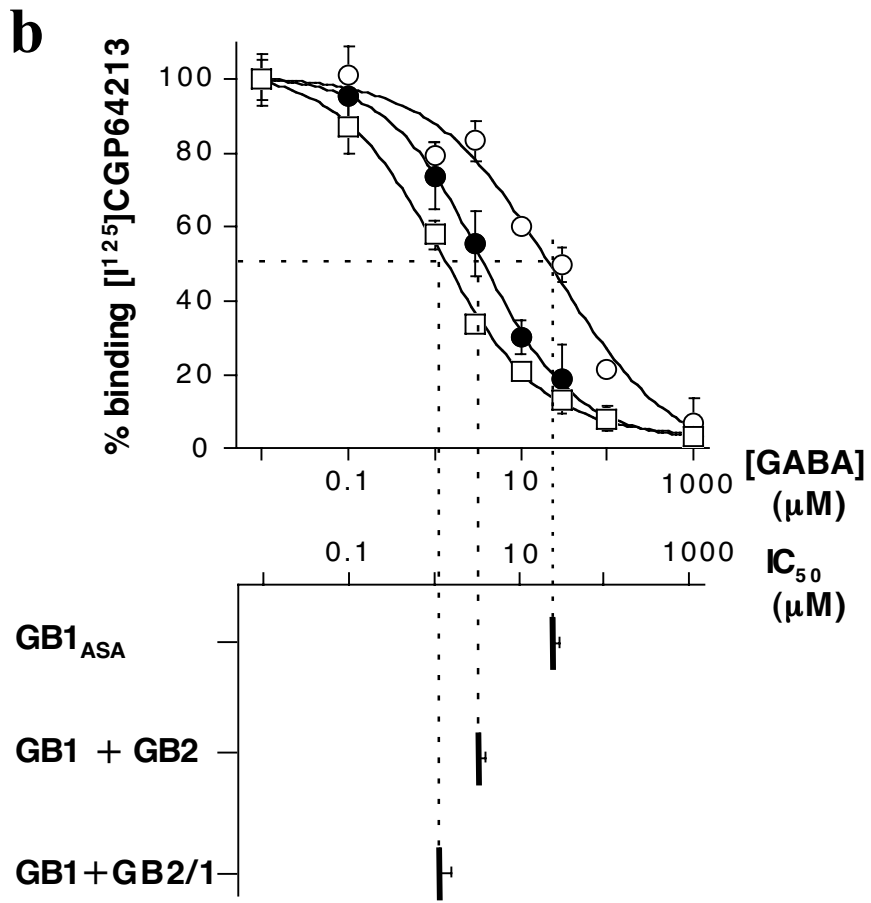

Figure 1

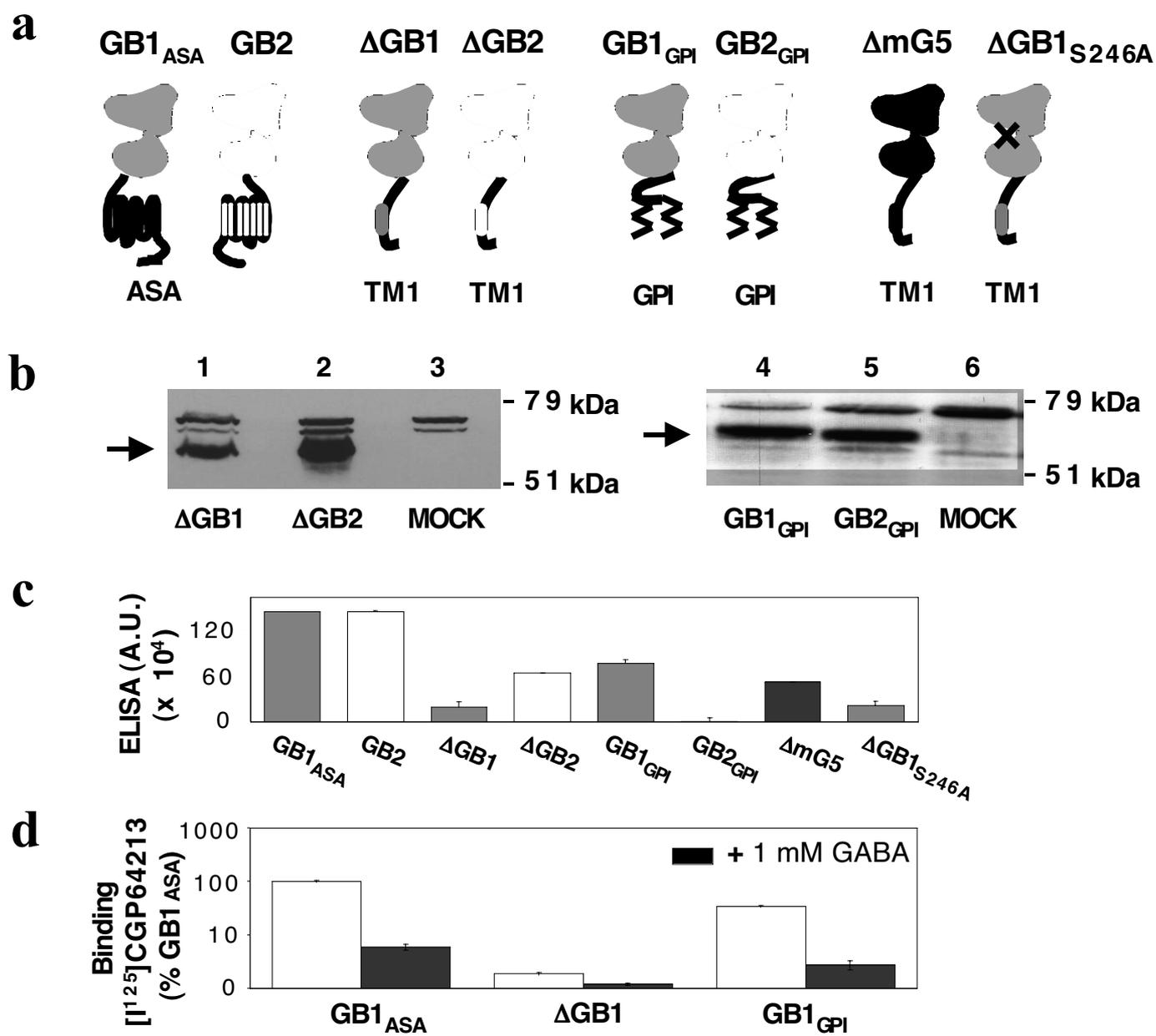

Figure 2

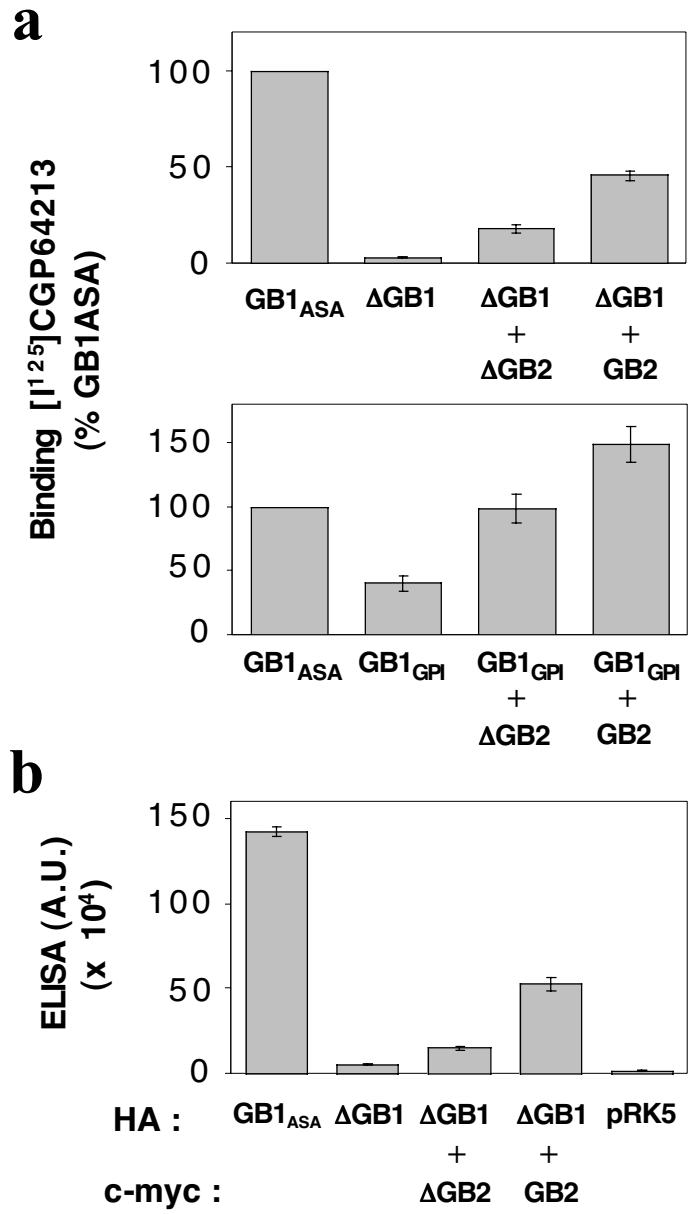

Figure 3

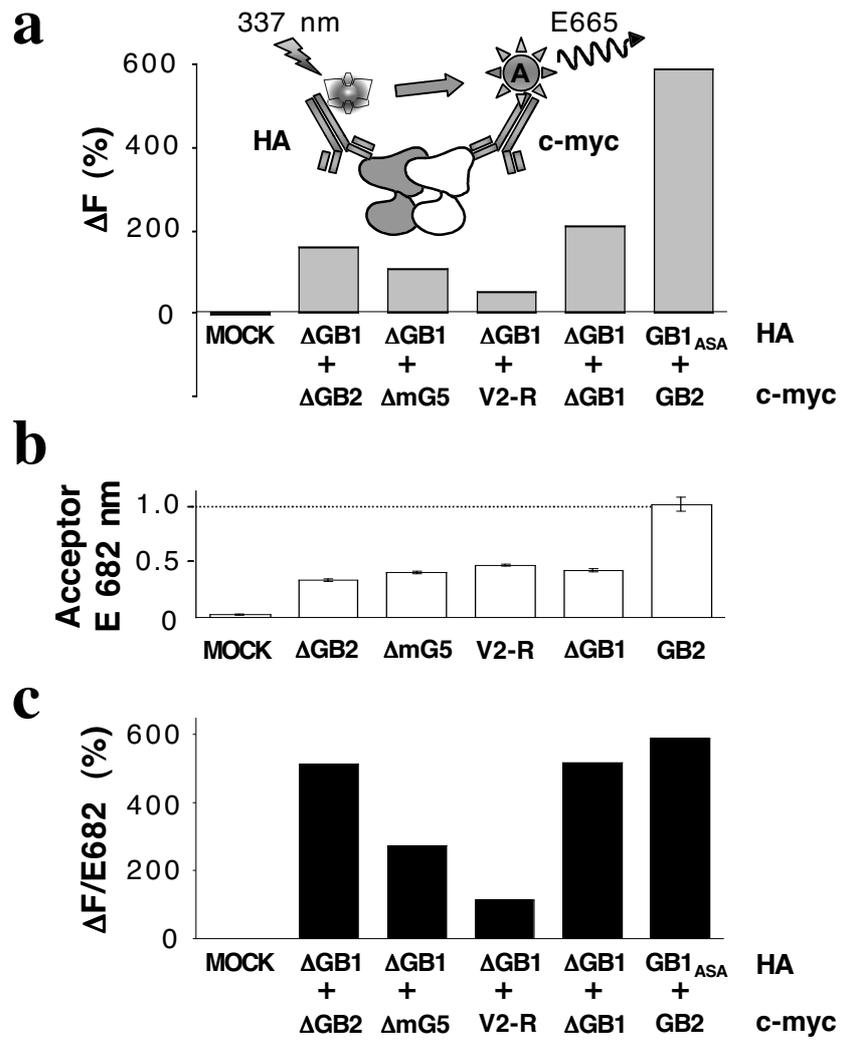

Figure 4

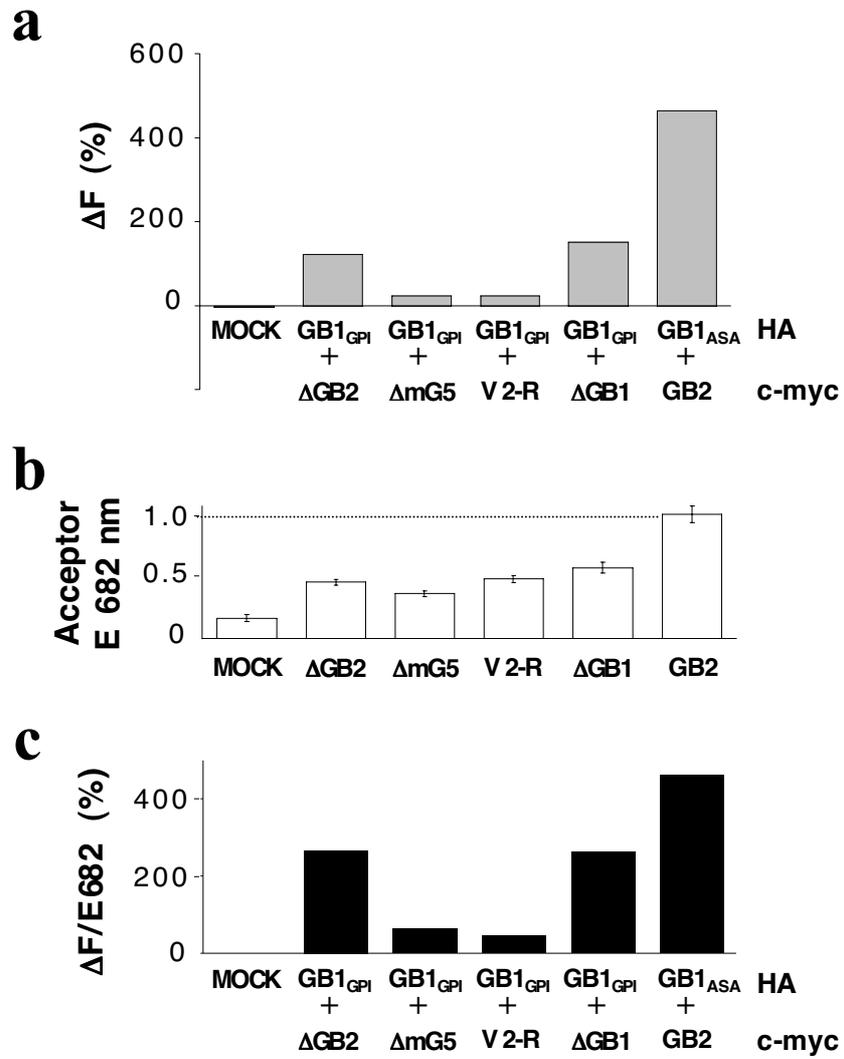

Figure 5

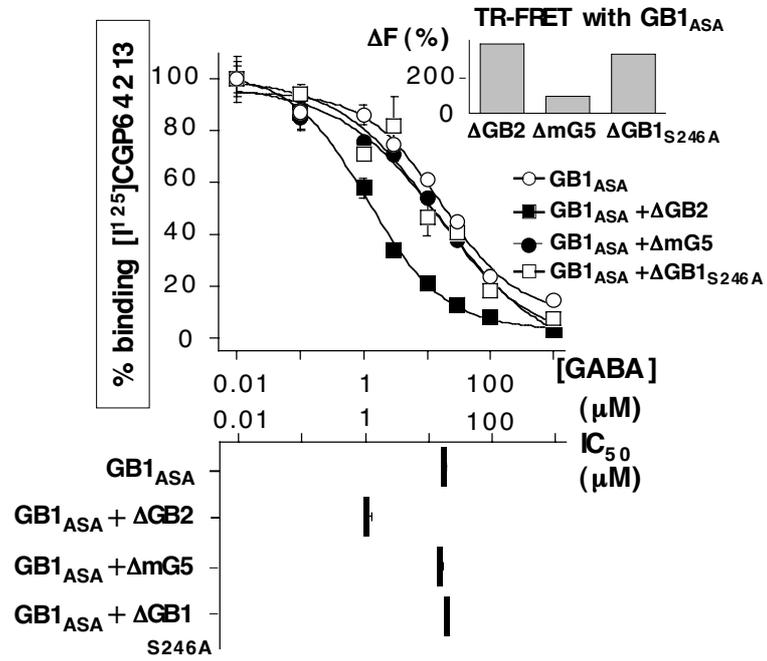
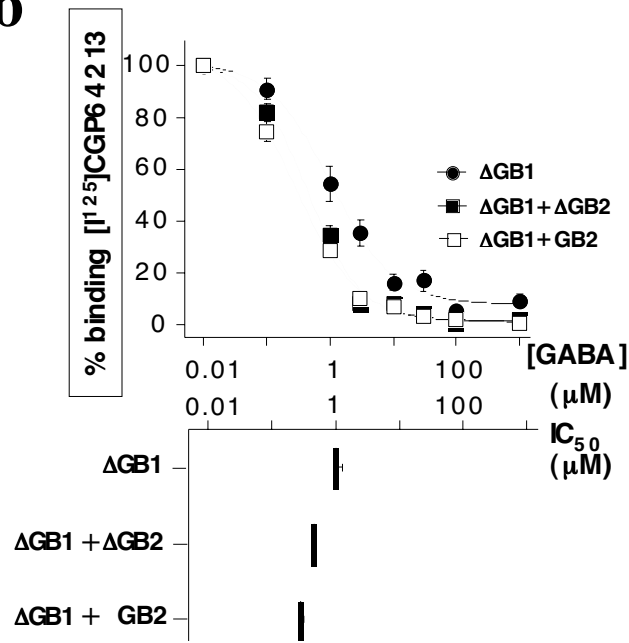

Figure 6